\documentclass[Physsubmission, Phys]{SciPost}

\hypersetup{
    colorlinks,
    linkcolor={red!50!black},
    citecolor={blue!50!black},
    urlcolor={blue!80!black}
}

\usepackage[bitstream-charter]{mathdesign}
\DeclareSymbolFont{usualmathcal}{OMS}{cmsy}{m}{n}
\DeclareSymbolFontAlphabet{\mathcal}{usualmathcal}

\begin{document}

\begin{center}{\Large \textbf{
Dark Matter Bound States:\\
A Window into the Early Universe 
}}\end{center}

\begin{center}
Juri Smirnov\textsuperscript{1},
\end{center}

\begin{center}
{\bf 1} Department of Mathematical Sciences, University of Liverpool, Liverpool, L69 7ZL, United Kingdom
\\
juri.smirnov@liverpool.ac.uk
\end{center}

\begin{center}
\today
\end{center}


\definecolor{palegray}{gray}{0.95}
\begin{center}
\colorbox{palegray}{
  \begin{tabular}{rr}
  \begin{minipage}{0.1\textwidth}
    \includegraphics[width=30mm]{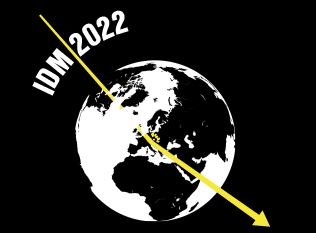}
  \end{minipage}
  &
  \begin{minipage}{0.85\textwidth}
    \begin{center}
    {\it 14th International Conference on Identification of Dark Matter}\\
    {\it Vienna, Austria, 18-22 July 2022} \\
    \doi{10.21468/SciPostPhysProc.?}\\
    \end{center}
  \end{minipage}
\end{tabular}
}
\end{center}

\section*{Abstract}
{\bf
Despite significant experimental sensitivity to point-like, weakly interacting particles at the electroweak mass scale, dark matter has not been found yet. This could hint at a more complex dark sector with multiple states or composite dark particles, much like the complexity of the standard model. Currently, our experimental sensitivity to such scenarios is limited by a lack of theoretical understanding. The investigation of Dark Matter systems with unstable or stable bound states provides a playground that leads to the development of important tools needed for the understanding of dark sectors with complex phenomena and allows to experimentally test well-motivated DM scenarios.}

\vspace{10pt}
\noindent\rule{\textwidth}{1pt}
\tableofcontents\thispagestyle{fancy}
\noindent\rule{\textwidth}{1pt}
\vspace{10pt}

\section{Introduction}
\label{sec:intro}
The particle nature of dark matter is one of the foremost, central problems of high-energy physics. But
despite overwhelming astrophysical evidence for its existence, key properties, such as its interaction strength, and its precise impact on structure formation remain unknown. Similarly, the production mechanism of dark matter is a longstanding, central problem of cosmology.  Importantly, in specific particle-physics scenarios, these two problems are tightly linked. Thus, experimental discovery and mass determination would
provide a new building block for particle physics, pointing toward new structures, symmetries, and
energy scales that lie beyond the standard model, while also providing deep new insights into how to model the evolution of the universe.

Why, despite unprecedented experimental sensitivity --- for example with LZ~\cite{LZ:2022ufs,Arcadi:2017kky}, and FermiLAT~\cite{Fermi-LAT:2015att} ---, have we failed to detect this new particle? The assumption that dark matter (DM) is a single, elementary particle with an electroweak scale mass (e.g., a vanilla WIMP, or weakly interacting massive particle) may be too simple. The dark sector could be complex, just like the standard model (SM). It could feature new strong dynamics, number-changing interactions, multiple stable dark states, and composite objects. In such scenarios, the stable relics of the dark sector would be produced very differently from the simplest thermal relics, and feature very different properties at present times, changing how we should search for them.  Even for TeV-scale WIMPs, self-interaction effects can lead to multiple bound states with severe effects on production and phenomenology.

\section{Unstable Dark Matter Bound States}

The WIMP remains one of the most compelling DM scenarios, and its parameter space above the GeV scale is wide open~\cite{Leane:2018kjk}. It would be negligent to not fully explore it experimentally. As discussed in Ref.~\cite{Mitridate:2017izz}, this classic scenario can be subject to complex phenomena, such as bound-state formation. This possibility is still under-explored, understanding those effects will strongly improve experimental sensitivity, and help attack the WIMP window from the high-mass end~\cite{Bottaro:2022one,Bottaro:2021snn}. 

Studies of a $U(1)$ toy mode provide an interesting example to understand in which situations bound-state formation during freezeout will become relevant ~\cite{vonHarling:2014kha,Cirelli:2016rnw,Kahlhoefer:2020rmg}. However, in a concrete model of weakly interacting massive particles (WIMPs), the situation can become more complex, especially in the case of non-Abelian force mediators.  An important group of publications investigated bound-state formation effects in the simplest models of WIMPs. For example, the electroweak triplet was studied in Refs.~ \cite{Braaten:2017gpq, Asadi:2016ybp}. Further studies of thermal effects in this framework have been performed in~\cite{Binder:2019erp, Harz:2019rro, Harz:2018csl, Baldes:2020hwx}.
 
In terms of searches, experiments such as the Fermi-LAT, the Cherenkov telescope H.E.S.S~\cite{HESS:2018kom} as well as the planned CTA~\cite{Morselli:2017ree}, and LHASSO~\cite{DiSciascio:2016rgi} experiments, will soon be dominating the search for thermally produced WIMP-like dark matter. Note that also neutrino telescopes, such as Antares and IceCube~\cite{ANTARES:2020leh} will gain in sensitivity and begin to place severe bounds on dark matter annihilation into neutrino final states.

A particular advantage of all those experiments is, that they rely on exactly the same process which is responsible for the thermal relic production in the early universe and thus provide the best way to address the WIMP-like thermal production hypothesis.  As will be discussed below the study of dark matter bound-state formation will supply the needed theoretical framework to allow the indirect detection experiments to progress faster into the open parameter space.

\begin{figure}[h]
\centering
\includegraphics[width=0.6\textwidth]{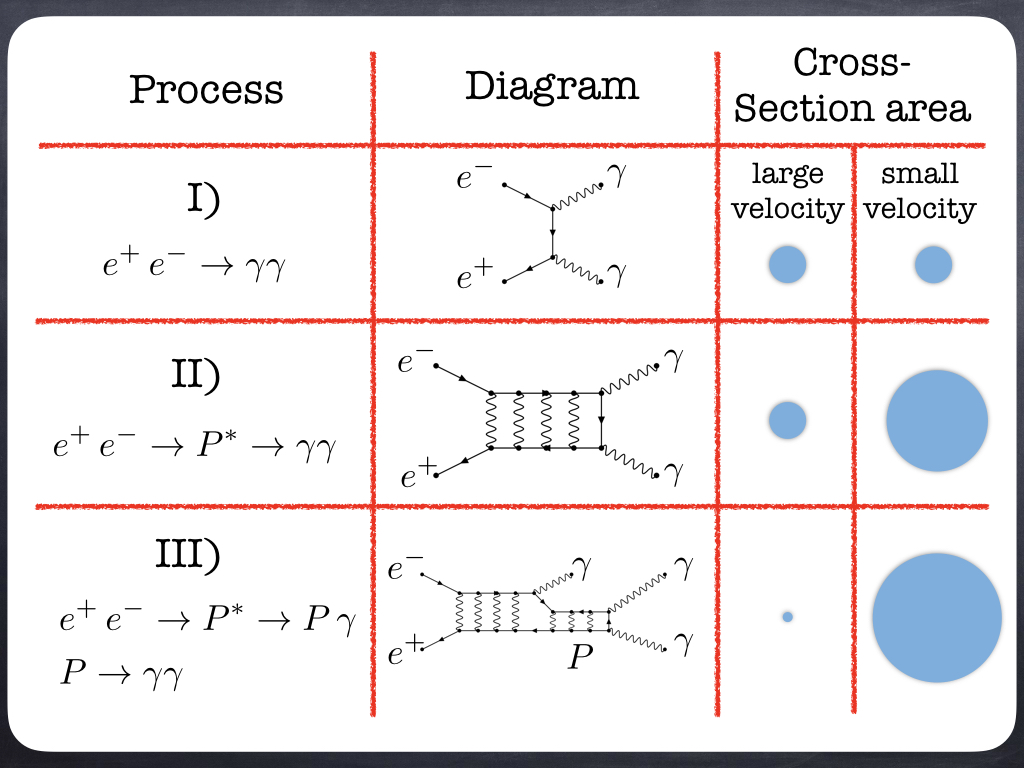}
\caption{Electron positron annihilation exemplifies non-perturbative effects at low velocities. Top to bottom: tree-level annihilation, annihilation with Sommerfeld enhancement, and annihilation via positronium formation. Circle size schematically depicts the relative size of the cross sections.}
\label{fig:bs}
\end{figure}
\subsection{Effects on the Dark Matter Abundance}

An important input for the experimental searches is the target annihilation reaction rate, provided by the production process. In a detailed study, important methodologies have been developed~\cite{Mitridate:2017izz}, that can be used to describe dark-matter bound-state formation for arbitrary gauge groups. The developed formalism relies on the non-relativistic expansion and the dipole approximation, which is well applicable to all weakly coupled models. Further studies have been performed by taking into account finite temperature effects within the framework of thermal quantum field theory~\cite{Binder:2021vfo,Binder:2020efn}.  

In Fig.~\ref{fig:bs}, I sketch the general idea of the bound-state effect on the dark matter freezeout, using electron-positron annihilation as an example. Here, it is well known, that at low velocities the non-relativistic potential between the particles will enhance the annihilation rate. Furthermore, the total rate is dominated by the on-shell formation of positronium, an $e^+ e^-$ bound state, that consequently decays to photons. The situation is similar in DM systems. Here, however, early universe conditions have to be taken into account. At large temperatures, the forming bound states are quickly broken by thermal collisions, but when temperatures fall, the decay of forming bound states dominates and provides an additional DM depletion channel. Thus on general grounds the total DM annihilation rate increases, which also increases the DM mass at which the correct relic abundance is reproduced. 

This effect on the freezeout process can be important in many scenarios. Two cases should be distinguished, one is the scenario where the mediating particle is massless, as is the case in color co-annihilation~\cite{Ellis:2015vna,Gross:2018zha}. The second case becomes relevant, where despite the fact that the potential has a finite range, bound states can be formed. This is the case when the Bohr radius of the forming bound state is smaller than the Yukawa radius, set by the mediator mass, a condition that can be formulated as $ M_V < \alpha M_{\rm DM}$, where $M_{\rm DM}$ is the DM mass, $\alpha$ the effective coupling strength and $M_V$ is the mass of the mediating particle. This condition is satisfied in the case of DM which carries an electroweak ($SU(2)_L$) charge. In particular, EW multiplets that are larger than the triplet are increasingly strongly affected~\cite{Smirnov:2019ngs,Bottaro:2021snn}.  A concrete example is the $5$, or quintuplet, representation of electroweak DM. This representation is particularly interesting since it is automatically stable without the need to postulate an additional symmetry~\cite{Cirelli:2005uq}. In this case, as we showed in Ref.~\cite{Mitridate:2017izz}, the expected DM mass is increased by almost a factor of two and is expected to be around $14$ TeV. I will discuss interesting phenomenological implications, using this example candidate in the next section.

\begin{figure}[h]
\centering
\includegraphics[width=0.6\textwidth]{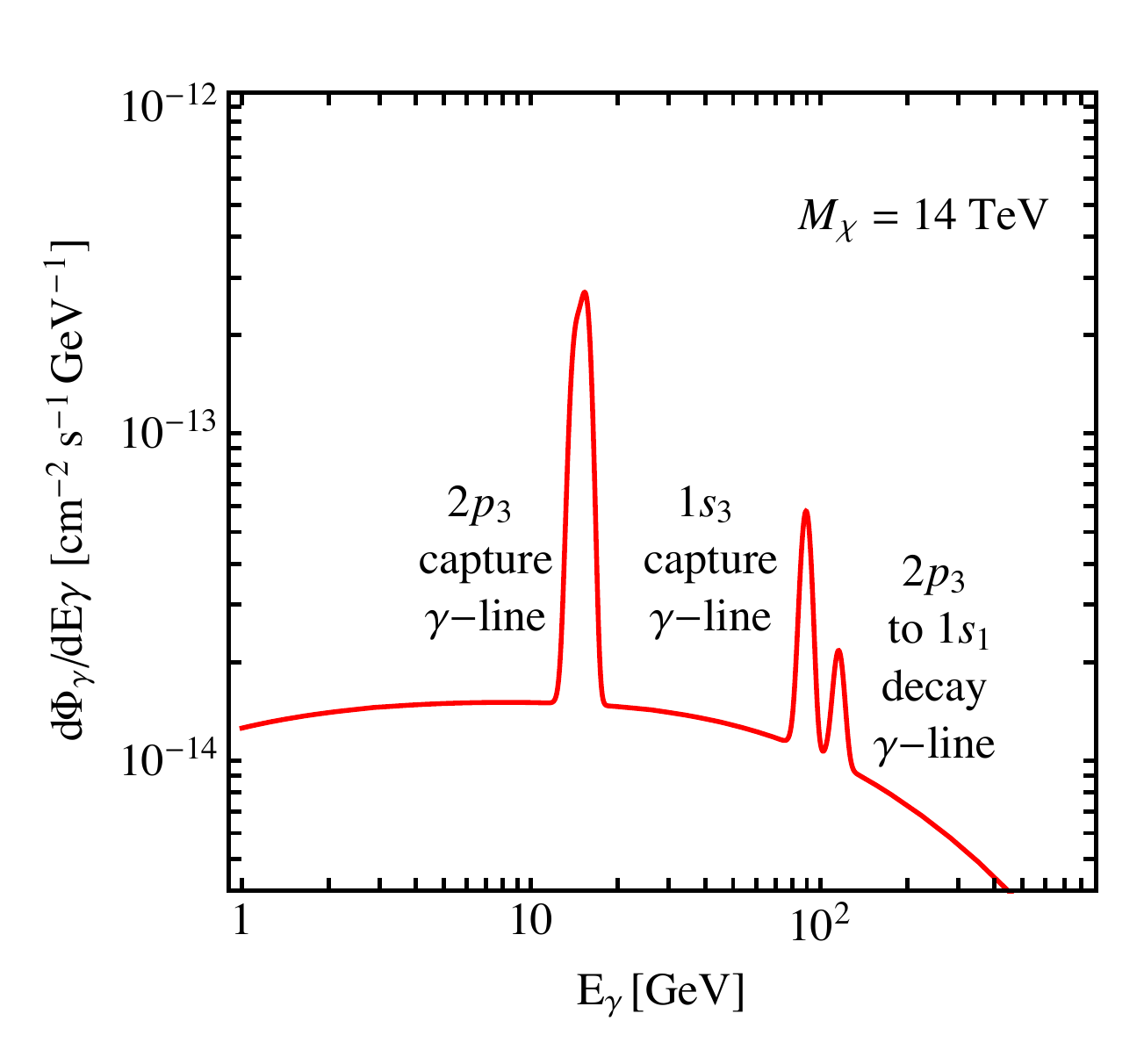}
\caption{The low-energy gamma-ray spectrum of an electroweak $5-$plet DM candidate, which shows the gamma-ray lines from capture photons on top of the continuum fragmentation spectrum based on Refs.~\cite{Smirnov:2019ngs}.}
\label{fig:lineflux}
\end{figure}

\subsection{New Signals from late-time Bound-State formation}

As known from SM observations, the formation of bound states is accompanied by particle emission to compensate for the released binding energy. Given that DM is an element of the electroweak multiplet photon emission is possible in the bound-state formation process, which provides an excellent, new signature for indirect DM searches. Typical binding energies scale as $E_B \sim \kappa \alpha^2 M_{\rm DM}$, where $\kappa<1$ is a factor that depends on the bound state quantum numbers. Thus the typically emitted energy of the capture-photon will be orders of magnitude lower than the DM mass. This allows probing very heavy DM candidates by studying photon emission at lower energies.

In Fig.~\ref{fig:lineflux}, I show the line spectra of bound-state formation processes in a model with a Majorana fermion charged under the $SU(2)_L$, which lives in the 5-element multiplet, the quintuplet~\cite{Mitridate:2017izz,Smirnov:2019ngs}. 
The relative line height and positions allow for a unique reconstruction of DM mass and gauge charge, or in other words its multiplicity under the electroweak gauge group. This is a unique situation, where an indirect DM search can provide exact particle physics properties of the DM candidate. This spectroscopic method will become particularly relevant for sensitivity improvement of gamma-ray experiments to heavy dark matter candidates. This method will greatly boost the sensitivity of H.E.S.S~\cite{HESS:2018kom},  LHASSO~\cite{DiSciascio:2016rgi}, as well as the planned CTA~\cite{Morselli:2017ree}, to high DM masses.


\section{Stable Dark Matter Bound States}
\label{sec:stable}

Building blocks of SM matter are known to be composite. In particular, the structure of the strong interaction that leads to composite nucleons has a conserved global symmetry, which is responsible for the long proton lifetime, the baryon number. 
It is plausible that the stability of DM particles originates from a similar principle, and DM is a stable composite object, see for example Refs.~\cite{Antipin:2015xia,Mitridate:2017oky}.  Interestingly, the simple assumption that DM is a composite object opens up the possibility that it is made of constituents charged under the strong nuclear force~\cite{DeLuca:2018mzn}, an option that has been previously deemed impossible. 

Changing to the simple assumption that DM is a composite particle, can significantly affect the production in the early universe and the detection prospects at late times. In particular, since DM which is composite necessarily features multiple states, including excitations, the understanding of its production process can provide much more insight into the conditions of the early universe. Given the high temperatures at which the product is expected to proceed, this provides a new window into the early universe, at unprecedentedly early times. 

\subsection{Freezeout in Confining Dark Sectors}

As a concrete example of a composite DM model, I will discuss a dark QCD type model, with the main difference being that the constituent dark quarks have masses well above the confinement sale. 
In the confined state, this sector has a stable particle, which is the lightest baryon, stabilized by global, dark baryon-number conservation. The dark mesons, carry no baryon number and are typically unstable. The lightest particles are glueballs, composite gluonic states, that are also unstable but can feature intermediate lifetimes, due to a suppressed decay operator.  
This scenario has been investigated in Refs.~\cite{Mitridate:2017oky,Dondi:2019olm}, and perturbative bound-state formation was included in the computation. However, the effects of the phase transition (PT), which in this case is a first-order PT, have not been incorporated. Recently, we demonstrated that indeed the PT effect changes the expected relic abundance by orders of magnitude. 

\begin{figure}[h]
\centering
\includegraphics[width=0.6\textwidth]{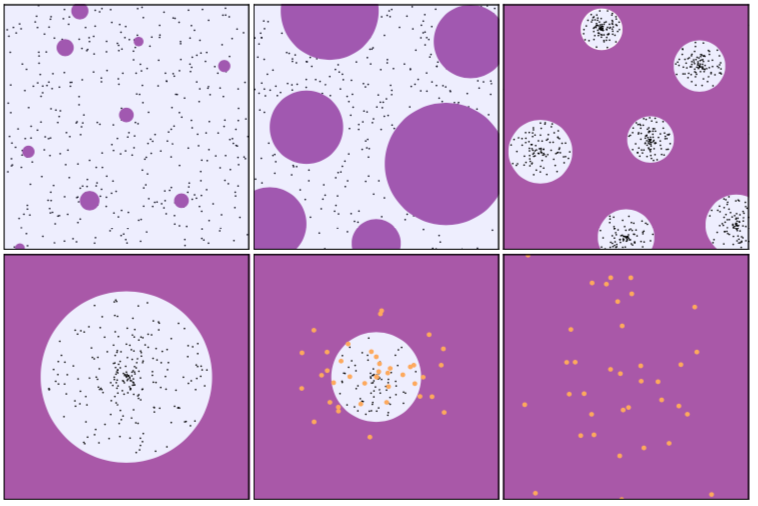}
\caption{Schematic representation of the first order PT and its effect on the dark baryon abundance. Top row: bubble nucleation, bubble growth, percolation, and the formation of pockets with trapped free quarks. Bottom row: shrinking of the pockets, recoupling of the bound-state formation interactions and baryon leakage from the pocket, confinement with a reduced population of color-neutral dark baryons.}
\label{fig:bubbles}
\end{figure}

In Fig.~\ref{fig:bubbles}, I show how a first-order phase transition can significantly affect the number of relic particles in the universe, by compressing them in shrinking false vacuum pockets. 
In a recent study~\cite{Asadi:2021pwo,Asadi:2021yml}, my collaborators and I demonstrated that in a large class of dark matter models with a confining phase transition this effect indeed strongly suppresses the dark-sector relic abundance. 
In light of our findings, many concrete studies of confining dark sectors need to be re-evaluated.

Furthermore, if light degrees of freedom with a relativity long life-time are present in the theory their decay can lead to further dilution of the DM states~\cite{Asadi:2021bxp}, as discussed in Ref.~\cite{Asadi:2022vkc} such degrees of freedom, the glueballs, are unavoidable in this type of theories, and in large fractions of the parameter space have lifetimes up to a second. This leads to an additional DM dilution and moves the expected mass scale to the multi PeV scale. This mass is orders of magnitude above the value expected for a thermal DM candidate that annihilates with a cross section that respects the unitarity bound~\cite{Griest:1989wd,Smirnov:2019ngs}.  
\subsection{New Experimental Signatures}

In previous work, we have proposed that metastable, excited DM bound states can lead to phenomena similar to radioactive decay in the dark sector~\cite{Mitridate:2017oky}. 
The important conclusion of this finding would be, that ultra-heavy thermal dark matter can be searched for in indirect detection experiments. This was previously deemed hopeless, as the low number density suppresses the annihilation signal. A signal that is due to decay of the dark state, which is metastable, scales linearly with the number density is thus much stronger, and allows probing even very heavy DM scenarios.

Another possibility to search for dark sector relics with large masses arises if the particles have large elastic cross sections, of the order of a barn, with standard model particles. Such strong interactions might have been missed by direct detection and other searches. In particular, the low background detectors, such as XENON1T would be blind to such strong interactions~\cite{Cappiello:2020lbk}. Since the dark sector particles would scatter multiple times when passing through the detector, the signals would be rejected as background. Implementing coincidence measurement techniques can open up the window on the strong interacting relics with little additional technical effort. In particular as liquid noble gas, and neutrino detectors gain in volume one would expect coincident signals to play a role, and dedicated analyses are being developed~\cite{Aalbers:2022dzr}.

Intriguingly, the compression mechanism, discussed above, opens up the possibility of forming compact macroscopic objects, see for example~\cite{Bottaro:2021aal,Kawana:2021tde}. Despite their low number density, their existence can lead to striking signatures, such as white dwarf star ignition~cite{Graham:2018efk}. Recent observations of a new class of supernovae events, the so called Calcium-rich gap transients, could result from such white dwarf-DM collisions~\cite{Smirnov:2022zip}. This would explain their low progenitor masses, which seems to be well below the Chandrasekhar limit.

Composite DM models can provide, on the one hand, large SM-DM cross sections, and on the other, feature final states of DM annihilation with lifetimes up to a second. This opens up two new avenues to search for DM using extrasolar planets and brown dwarfs. One search is based on energy deposition and heating of the objects~\cite{Leane:2020wob}. The other possibility is to search for annihilation products of the secondary decays of dark sector particles with intermediate lifetimes~\cite{Leane:2021ihh}. Both searches are significantly affected by DM distribution in the celestial objects, and in particular populations of captured DM close to the surface can lead to striking signatures~\cite{Leane:2022hkk}. 
The research area of extrasolar planets is rapidly growing. The new infrared telescope JWST has already started data collection~\cite{2022arXiv220900620M}, and with other infrared observatories that will be launched in the near future, we expect excellent data in the coming few years.

\section{Conclusion}
I have discussed how a simple change in our assumptions about dark matter substrucutre, can have profound implications on the properties of the dark sector. In particular, leaving behind the single particle paradigm can lead to a very different freezeout phenomenology, and thermally produced dark matter states with masses well above the unitarity bound. Finally, the dark sector bound-state phenomena allow for investigating those scenarios in a variety of experimental searches, many of which have a great short-term optimization potential. 

\section*{Acknowledgements}
I thank the organizers of the iDM 2022 conference and in particular Dr. Valentyna Mokina, for organizing this community effort for preserving the discussions of iDM 2022 in form of this proceeding series. 

\paragraph{Funding information}
I acknowledge funding by the ERC under grant Number 742104.

\bibliography{SciPost_Smirnov_BiBTeX_File.bib}

\nolinenumbers

\end{document}